\documentclass[natbib]{apa6}
\usepackage[english]{babel}
\usepackage{inputenc}
\usepackage{amsmath}
\usepackage{graphicx}
\usepackage[colorinlistoftodos]{todonotes}
\usepackage{booktabs}
\usepackage{multirow}
\usepackage{apacite}
\usepackage[flushleft]{threeparttable}
\usepackage{url}
\graphicspath{{../Figures/}}

\newrobustcmd{\disambiguate}[3]{#3}

\title{Network-based statistics distinguish anomic and Broca aphasia}
\shorttitle{Anomic and Broca's aphasia}

\author{Xingpei Zhao$^1$, Nicholas Riccardi$^2$, Dirk-Bart den Ouden$^3$, Rutvik H. Desai$^2$, Julius Fridriksson$^3$, Yuan Wang$^{1*}$}
\affiliation{$^1$Department of Epidemiology and Biostatistics, University of South Carolina\\
         $^2$Department of Psychology, University of South Carolina\\
	$^3$Department of Communication Sciences and Disorders, University of South Carolina\\$*$Correspondence author: Department of Epidemiology and Biostatistics, Discovery I, 915 Greene Street, Columbia, South Carolina, U.S.A.. Email: wang578@mailbox.sc.edu.}
\leftheader{Zhao, Desai, Den Ouden, Fridriksson, Wang}

\abstract{Aphasia is a speech-language impairment commonly caused by damage to the left hemisphere. Due to the complexity of speech-language processing, the neural mechanisms that underpin various symptoms between different types of aphasia are still not fully understood. We used the network-based statistic method to identify distinct subnetwork(s) of connections differentiating the resting-state functional networks of the anomic and Broca groups. We identified one such subnetwork that mainly involved the brain regions in the premotor, primary motor, primary auditory, and primary sensory cortices in both hemispheres. The majority of connections in the subnetwork were weaker in the Broca group than the anomic group. The network properties of the subnetwork were examined through complex network measures, which indicated that the regions in the superior temporal gyrus and auditory cortex bilaterally exhibit intensive interaction, and primary motor, premotor and primary sensory cortices in the left hemisphere play an important role in information flow and overall communication efficiency. These findings underlied articulatory difficulties and reduced repetition performance in Broca aphasia, which are rarely observed in anomic aphasia. This research provides novel findings into the resting-state brain network differences between groups of individuals with anomic and Broca aphasia. We identified a subnetwork of, rather than isolated, connections that statistically differentiate the resting-state brain networks of the two groups, in comparison with standard lesion symptom mapping results that yield isolated connections.}


\begin{document}
\maketitle

\section{Introduction}
Aphasia is a speech-language disorder that commonly develops after a left-hemisphere stroke and can affect a person’s ability to read, understand, and speak a language. According to the National Aphasia Association, aphasia currently affects over two million Americans. 
There are several types of aphasia based on individual behavioral symptoms. Classification of aphasia types remains controversial.  Individuals with different types of aphasia often exhibit overlapping set of symptoms, while those within the same type can differ significantly. Nonetheless, some broad patterns associated with aphasia types can be observed. 

Two common types of post-stroke aphasia that are primarily characterized by speech production problems, as opposed to comprehension problems, are anomic and Broca's aphasia. Anomic aphasia is a fluent type of aphasia characterized by problems with word retrieval. Individuals with anomic aphasia may block upon failing word retrieval, or use circumlocution, but their speech fluency, repetition, comprehension, and grammatical speech are relatively preserved \citep{Dronkers2009}. On the other hand, Broca's aphasia is a non-fluent type of aphasia that may exhibit agrammatic speech output and poor repetition \citep{Hickok2009}. Their comprehension of spoken and written language often remains intact or has relatively mild impairments. As difficulty with word retrieval is related to damage throughout much of the peri-Sylvian region \citep{Dronkers2009}, word finding difficulties can be observed in both anomic and Broca's aphasia \citep{Whitaker2007}. However, the severity level and performance patterns are different in the two types of aphasia. For instance, it has been reported to be more difficult for individuals with anomic aphasia to retrieve nouns and for individuals with Broca’s aphasia to retrieve verbs \citep{Miceli1988, Zingeser1990}. Behaviorally and diagnostically, the primary difference between speakers with Broca’s aphasia and anomic aphasia is the typically reduced speech fluency and repetition performance in Broca’s aphasia \citep{kertesz2007}. In addition, Broca’s aphasia often co-occurs with expressive agrammatism, in the form of reduced sentence complexity and problems with morphosyntactic markers, such as inflections and function words, in production \citep{Friedman2009, Goodglass2001, Hillis2007}. It must be noted, however, that neither the articulatory deficit nor the agrammatism are essential diagnostic features of the Broca’s aphasia syndrome.

In terms of neural mechanisms, the left inferior frontal regions, especially pars triangular and pars opercularis traditionally referred to as 'Broca's area' (although see \citep{Tremblay2016} regarding terminology), have traditionally been associated with syntactic processing in language production and comprehension \citep{Caramazza1978, Keller2009}. The impairment to Broca's area alone, however, does not explain impaired syntactic comprehension and production in Broca's aphasia \citep{Mohr1978}. Grammatical impairments are associated with damage to a larger area surrounding Broca's area or other different brain area, such as lateral portion of the anterior temporal lobe \citep{Dronkers2009} and even posterior temporal cortex \citep{denOuden2019}. One hypothesis is that the impairment in a collection of speech-language functions in post-stroke Broca's aphasia is related to the damage to brain regions supplied by the superior division of the left middle cerebral artery \citep{Dronkers2009, Hillis2007, Dronkers2004, Zaidel1995}. Besides, grammatical and sentence-level language processing is a higher-lever language function and requires a series of complex mechanisms to process multiple information sources \citep{Caplan2000, Fiebach2006, Matchin2020}. As such, dysfunction within intrinsic brain networks may explain the agrammatic speech and difficulties in comprehension of complex sentences often observed in Broca's aphasia \citep{Zhu2014, Tomasi2012}. On the other hand, the object-naming difficulty in anomic aphasia cannot typically be attributed to damage in specific brain regions \citep{Yourganov2015,Fridriksson2018}. Interactions among the brain areas and circuits involved in sentences and grammatical-level processing in anomic and Broca's aphasia are not fully understood. 

Functional magnetic resonance imaging (fMRI) \citep{Ogawa1992, Rogers2001} is a neuroimaging technique for  investigating functional brain activity through blood-oxygen-level-dependent (BOLD) fluctuations in different brain regions. Network models can be applied to examine inter-relationships of these fluctuations in regions of interest (ROIs). In a brain network built on ROI-segmented fMRI data, the ROIs serve as nodes and functional connections between the ROIs serve as edges. The strength of functional connections is typically measured by correlation between the BOLD signals. Resting-state fMRI (rs-fMRI) is acquired in the absence of any tasks \citep{Fox2007}. Brain network models based on rs-fMRI describe intrinsic coherent functional activity in a resting brain and are considered to be permanent, trait-like functional signatures \citep{Hjelmervik2014}.  

Studies comparing resting-state networks from stroke-survivors with aphasia and healthy controls have revealed that the peri-Sylvian region, posterior middle temporal gyrus, anterior superior temporal gyrus, superior temporal sulcus, and Brodmann's area 47 of the inferior frontal gyrus in the left hemisphere are involved in semantic processing and language comprehension \citep{Dronkers2004, Turken2011, Zhu2014}. These studies suggest that rs-fMRI is relevant in understanding network-level brain dysfunctions in people with aphasia. Their approaches focused on functional connectivity among selected ROIs in the left hemisphere. However, according to \citet{Yang2016,Yang2017}, intrinsic regional brain dysfunctions in aphasia are associated with functional connectivity patterns in regions across the brain. Consequently, it is essential to investigate whole-brain functional connectivity in aphasia. So far, only a few studies have assessed resting-state whole-brain functional connectivity 
\citep{Yang2016, Siegel2016, Guo2019, Yang2017}.  \citet{Siegel2016} predicted impairment in multiple behavioural domains based on associated resting-state functional connectivity and lesion location via machine-learning models. \citet{Yang2017} used multivariate pattern analysis to identify whole-brain functional connectivity patterns that distinguish stroke survivors with aphasia and healthy controls. To the best of our knowledge, no study has directly compared resting-state whole-brain functional connectivity between speakers with anomic versus Broca's aphasia. 

To identify distinct functional connections between two groups, a standard approach is to examine differences at each connection through mass univariate testing, where a large number of hypothesis tests are typically performed. After rs-fMRI data are preprocessed with an atlas, the number of ROIs can exceed 300. For instance, 384 ROIs are created by the atlas of intrinsic connectivity of homotopic areas (AICHA) \citep{Joliot2015}, yielding more than 70,000 connections between the ROIs. With such a large number of hypothesis tests, the adjusted $p$-values after standard multiple comparison correction are large and may fail to detect significant differences in connections. Thus, the mass univariate testing approach may provide insufficient power, especially when distinct connections are not independent. The network-based statistic (NBS) method \citep{Zalesky2010} is a graph-theoretic approach that provides an efficient process to identify a subnetwork that distinguishes two groups of brain networks. The NBS method offers a substantial gain in power when such a subnetwork exists. From the comparisons of networks between healthy controls and participants with schizophrenia \citep{Zalesky2010}, Alzheimer's Disease \citep{Zhan2016}, internet addiction \citep{Wen2016}, and borderline personality disorder \citep{Xu2016}, the NBS method has identified one or more altered subnetworks related to these neurological or psychiatric disorders. Furthermore, we can assess the network properties of the subnetwork by complex network measures \citep{Rubinov2010}. 

In this study, we apply the NBS method for the first time to rs-fMRI data of individuals with anomic and Broca's aphasia. We aim to identify a subnetwork that distinguishes the anomic and Broca's brain networks in the resting state, and to understand the role of the subnetwork in deficits of speech-language processing through its network properties. We also compare findings with standard voxel-, region- and connectivity-based lesion symptom mapping (LSM). 

\section{Methods}
\subsection{Study description}
Participants were recruited from the local community in Columbia, South Carolina, as part of the Predicting Outcome of Language Recovery in Aphasia (POLAR) study of post-stroke aphasia. Only participants with a single ischemic or a hemorrhagic stroke in the left hemisphere were included. The participants with lacunar infarcts or with isolated damage in brainstem or cerebellum were excluded. The research was approved by the Institutional Review Board (IRB) at the University of South Carolina.    

\subsection{Participants}

\begin{figure}[b!]
  \includegraphics[width = 1\linewidth]{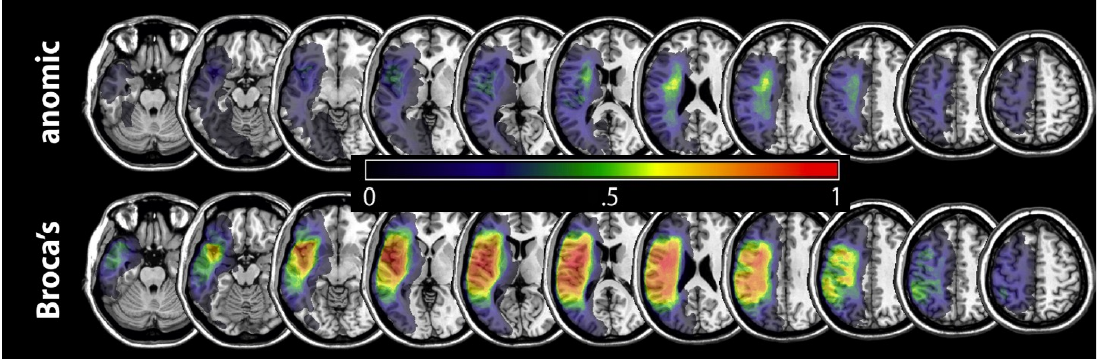}
  \caption{\label{fig: lesion}Overlap of lesions across patients of anomic and Broca’s aphasia. A voxel with overlap = 1 indicates that this voxel is lesioned in all patients.}
\end{figure}

Aphasia types were classified based on the Western Aphasia Battery-Revised (WAB-R) \citep{kertesz2007}. Among the 49 participants included in the study sample, 15 were diagnosed with anomic aphasia, and 34 were diagnosed with Broca's aphasia. Figure~\ref{fig: lesion} shows a lesion overlap map of the participants in the two groups. Demographic statistics of the two groups are summarized in Table \ref{demographic statistics}. The mean age in the anomic group was 62.73 y.o. (s.d. $=11.97$; range $=41$) and the mean age in the Broca's group is 59.82 y.o. (s.d. $=10.35$; range $=39$). Respectively $60\%$ and $68\%$ of the participants in the anomic and Broca's group were male. There was 
no significant difference in age and gender between the anomic and Broca's group (age: $p$-value $=0.42$ by two-sample $t$-test; gender: $p =0.74$ by $\chi^2$-test). The mean WAB-R score for the anomic group was $85.74$ (s.d. $=6.38$; range $=22.1$) and $46.44$ (s.d. $=16.93$; range $=59.1$) for the Broca's group. The WAB-R score for the anomic group is significantly higher than the Broca's group ($p$-value $<0.01$ by two-sample $t$-test), since the participants with Broca's aphasia tend to have a lower score in the section of fluency and repetition than the participants with anomic aphasia. 

\begin{table}[t!]
	\centering
		\begin{tabular}{lccc}
			& \textbf{Anomic} & \textbf{Broca's} & \textbf{$p$*}\\
			& $n_1=15$ & $n_2=34$ & \\
			\hline
			Age	&	62.73$\pm$11.97	&	59.82$\pm$10.35 &	0.42\\
			Race/Ethnicity &		&	 &	0.51 \\
			\hspace{2pt} Hispanic & 0 & 0 &\\
			\hspace{2pt} Black & 13 & 29 &\\
			\hspace{2pt} White &  87 & 68 &\\
			\hspace{2pt} Other &  0 & 3 &\\
			Education & &  & 0.26\\
			\hspace{2pt} High school  & 7 & 26 &\\
			\hspace{2pt} College or associate & 7 & 9 &\\
			\hspace{2pt} Bachelor and above  & 86 & 65 &\\
			
			Gender & & & 0.74 \\
			\hspace{2pt} Male  & 60 & 68& \\
			WAB-R		& 85.74$\pm$6.38 & 46.44$\pm$16.93 & $<0.01$\\
			\bottomrule
		\end{tabular}
		\caption{\label{demographic statistics}Demographic statistics of participants with anomic and Broca's aphasia in the POLAR study: mean and standard deviation of age and WAB-R score, percentages of race/ethnicity and education levels out of all participants in the study. (*Two-sample $t$-test was used to test group difference in age and WAB-R score; $\chi^2$-test and Fisher's exact test were used to test group difference in race/ethnicity, gender and education.)}
\end{table}

\begin{figure*}[t!]
	\centering
	\includegraphics[width=\linewidth]{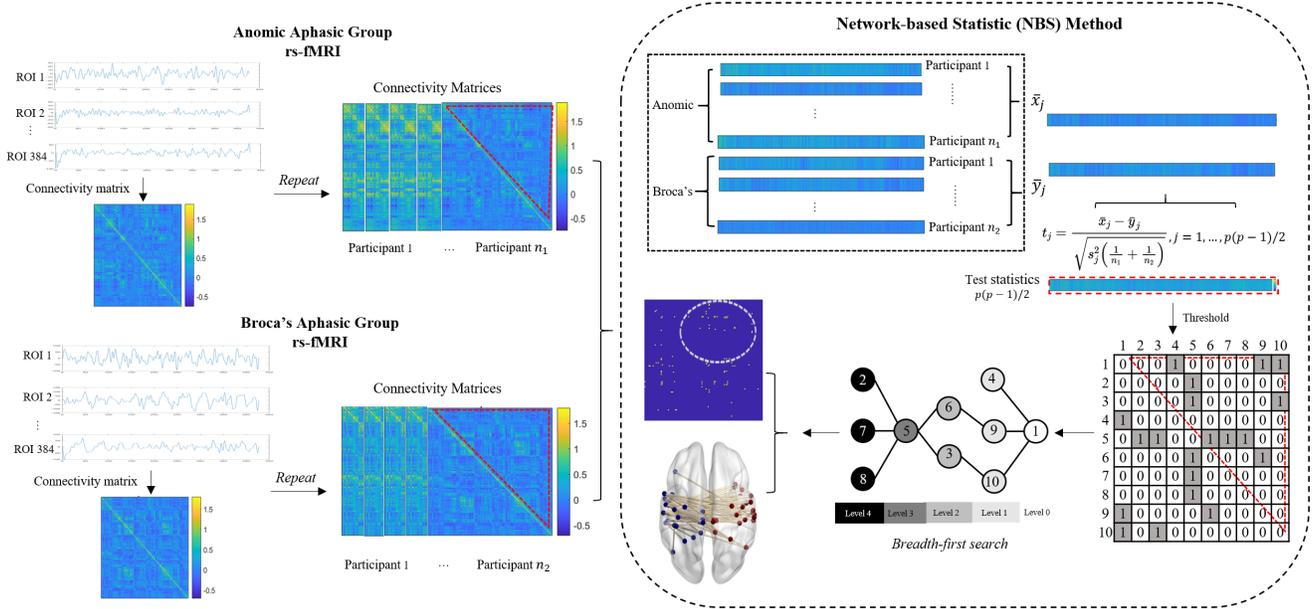}
	\caption{Left: Resting-state connectivity matrices are constructed from two groups of rs-fMRI data. Right: Illustration of the network-based statistic (NBS) method and how a connected component or subnetwork differentiating the connectivity patterns of the two groups of resting-state functional brain networks is identified through the breadth-first search algorithm.}
	\label{fig:Illustration of NBS}
\end{figure*}

\subsection{Data acquisition and preprocessing}
 The rs-fMRI data were acquired on a Siemens Prisma 3T scanner with a 20-channel head coil located at the Center for the Study of Aphasia Recovery at the University of South Carolina. The following imaging parameters of images were used: a multiband sequence (x2) with a $216 \times216$ mm field of view, a $90 \times 90$ matrix size, and a 72-degree flip angle, 50 axial slices (2 mm thick with $20 \% $ gap yielding 2.4 mm between slice centers), repetition time TR =1650 ms, TE=35 ms, GRAPPA=2, 44 reference lines, interleaved ascending slice order \citep{Grigori2018}. During the scanning process, the participants were instructed to stay still with eyes closed. A total of 370 volumes were acquired. The preprocessing procedures of the fMRI data include motion correction, brain extraction and time correction. By applying the atlas of intrinsic connectivity of homotopic areas (AICHA) \citep{Joliot2015}, 384 regions of interest (ROIs) were created. 

\subsection{Network-based statistic (NBS) analysis}

We used the NBS method to identify a subnetwork that differentiates connectivity patterns of the anomic and Broca's aphasia groups. An illustration of the analytical pipeline is summarized in Figure~\ref{fig:Illustration of NBS}. It builds on edge-level mass univariate testing with multiple comparison adjustment but refines the approach through a breadth-first search algorithm.

We first constructed functional connectivity matrices based on Pearson's correlation between BOLD signals in the ROIs. The Fisher's $z$-transformation was applied to the coefficients to enforce normality: 
\begin{equation}\label{fishertrans}
	z=\frac{1}{2}ln\left(\frac{1+r}{1-r}\right)=arctanh\left(r\right),
\end{equation}
where $r$ was a correlation coefficient and $z$ was the corresponding normalized correlation coefficient. After normalizing the coefficients, a $p \times p$ connectivity matrix, with $p = 384$ being the number of ROIs in the AICHA atlas, was constructed for each participant in the two aphasia groups. The $(i,j)$-th entry of such a matrix denotes the weight of an edge or connection between the $i$th and $j$th ROIs of the corresponding network. As the connectivity matrices are symmetric, we only included the $p(p-1)/2$ upper or lower triangular entries in the analysis.

To identify differences in the resting-state connectivity of the anomic and Broca's groups, we tested the null hypothesis 
$$
H_0: \mu_{1j} = \mu_{2j}, \text{ for all } j = 1,2,\dots, p(p-1)/2\text{,}
$$
where $\mu_{1j}$ and $\mu_{2j}$ are the respective $j$-th population means of the vectorized $p(p-1)/2$ edge weights. That is, we assumed no difference in the resting-state connectivity between the two groups under the null hypothesis. Under the alternative hypothesis, we assumed that at least one pair of mean edge weights are different between the two groups, i.e.
$$
H_1: \mu_{1j} \neq \mu_{2j}, \text{ for some } j = 1,2,\dots, p(p-1)/2 \text{.}
$$
The test statistic for comparing the $j$-th edge weights is 
\begin{equation}\label{teststatistic}
	t_j = \frac{\bar{x}_j  - \bar{y}_j}{\sqrt{{s_j}^2 \left(\frac{1}{n_1}+\frac{1}{n_2}\right)}} \text{,}
\end{equation}
where 
\begin{equation*}
	s_j^2 = \frac{(n_1-1)s_{1j} ^2+(n_2-1)s_{2j} ^2}{n_1+n_2-2} , 
\end{equation*}
with $\bar{x}_j$ and $\bar{y}_j$ being the respective $j$th sample means of the $p(p-1)/2$ edge weights, $s_{1j}$ and $s_{2j}$ being the respective standard deviations, and $n_1$ and $n_2$ being the respective sample sizes of the two groups. In total, $p(p-1)/2$ test statistics were computed. 

We then obtained a binary adjacency matrix by applying a chosen threshold to the edge-level test statistics. If a test statistic exceeded the threshold, the corresponding edge weight of the adjacency matrix was set to $1$ and otherwise 0. The edges that remained after the thresholding are called the {\em suprathreshold edges}. Connected component(s) or subnetwork(s) (there may be more than one), where all nodes in each subnetwork were linked by suprathreshold edges, were identified from the adjacency matrix through the breadth-first search algorithm \citep{Ahuja1993, Hopcroft1973}. An illustration of the algorithm is included in Figure \ref{fig:Illustration of NBS}. 

Similar to cluster-based thresholding of statistical parametric maps \citep{Nichols2002, Hayasaka2004}, we conducted a random permutation test to determine the statistical significance of the subnetwork(s) and control the family-wise error rate (FWER). In each permutation, participants in the two groups were shuffled and randomly divided into two new groups. Test statistics were computed by \eqref{teststatistic} based on the new grouping. The threshold imposed on the test statistics was the same for all permutations. Denoting the maximum of all possible subnetwork sizes in a permutation as $S^{*}$, the $p$-value for any observed subnetwork of size $S$ was calculated by $Pr\left( S^{*} \ge S \right)$, depicting how likely the size of the largest subnetwork in a permutation exceeds the size of the observed subnetwork. When the $p$-value was less than a certain significance level, which is set to $5\%$ in this study, we concluded that this subnetwork is significantly different between the two aphasia groups. 

For baseline comparison with the NBS method, mass univariate permutation testing with multiple comparison was performed. The $p$-value for comparing the $j$th edge weights with \eqref{teststatistic} was calculated by $Pr\left(|t^{*}_j| \ge |t_j|\right)$, where $t^{*}_j$ and $t_j$ were the test statistics computed from permuted data and original data respectively. This $p$-value describes how likely the absolute value of test statistic from permuted data exceeds the absolute value of test statistic from original data. Then, the following multiple comparison procedures were applied to correct multiple $p$-values and control the FWER: Bonferroni correction \citep{Bonferroni1936}, Holm's Bonferroni correction \citep{Holm1979}, and false discovery rate (FDR) control \citep{Benjamini1995}.

\begin{figure*}[t!]
	\centering
	\includegraphics[width=\linewidth]{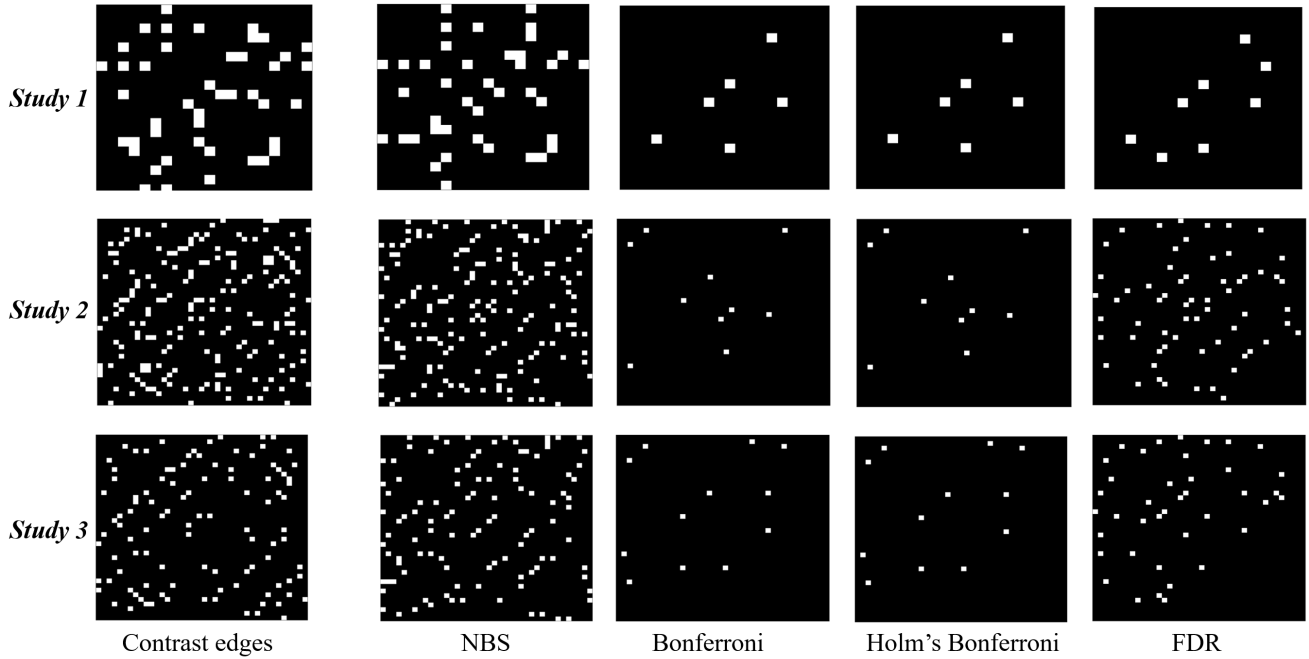}
	\caption{Simulation setup: The white blocks in the left side's matrices represent contrast edges, whose weights differ between the two groups. The white blocks in the right side's matrices are the subnetwork or connected component identified by the NBS method or the edges declared significant by multiple comparison.}
	\label{fig:Simulation illustration example}
\end{figure*}

\subsection{Performance evaluation of NBS}

We used three simulation studies to evaluate the empirical performance of the NBS method against  mass univariate testing with multiple comparison. We assessed two aspects of the performance: (1) sensitivity or true positive rate (TPR): the proportion of connections or edges containing group differences that are correctly identified; (2) $1-$ specificity or false positive rate (FPR): the proportion of edges without differences that are misclassified. Ideally,  TPR $=1$ (all edges that differ between the two groups are identified), and FPR $=0$ (all edges that do not differ between the two groups are not identified). Suppose $H$ is the set of edges that differ between the groups, $R$ is the set of edges that do not differ between the groups, and $\hat{h}$ is the set of edges comprising the subnetwork identified by a specific method (NBS, baseline mass univariate testing with multiple comparison). The TPR was then calculated by $|H\cap \hat{h}|/|H|$ and the FPR by $|R\cap \hat{h}|/|R|$.

In each of the three studies, we generated two groups of $p$-node networks. The group sizes are $n_1=n_2=10$ for all three studies. In each network, the weight of the edge between node $i$ and $j$ was generated by $w_{ij} \sim N(arctanh(r_{ij}), \sigma_{w}^2)$ with $r_{ij} \sim U(-1,1)$ and $\sigma_{w} = 1/\sqrt{p(p-1)/2-3}$. We randomly chose $C_r \%$ of $p(p-1)/2$ edges to differ in weights between Group 1 and 2, and refer to these edges as {\em contrast edges}. The weights of contrast edges were generated independently with 
	\[
w_{ij}^{*}=\left\{
\begin{array}{ll}
	w_{ij} + w^*, \text{for } w_{ij} \ge 0\\
	w_{ij} - w^*, \text{for } w_{ij} < 0 \text{,}
\end{array}
\right.
\]
where $w^{*}\sim N(0.03, 0.01)$. We compared the performance of the methods via different $p$ and $C_r$ values in the three studies.
\begin{description}
\item [\textbf{Study 1.}] $p=20$ and $C_r \% = 10\%$. 
\item [\textbf{Study 2.}] $p=40$ and $C_r \% = 10\%$. 
\item [\textbf{Study 3.}] 
$p=40$ and $C_r \% = 5\%$.
\end{description}

After the networks were generated, NBS with threshold 2.5
and mass univariate testing with multiple comparison were performed (Figure \ref{fig:Simulation illustration example}). We repeated the simulation process 5,000 times for each study. Average TPR and FPP were computed respectively. 

\subsection{Properties of subnetwork identified through NBS}

\begin{figure*}[t!]
\centering
  \includegraphics[width = 0.65\linewidth]{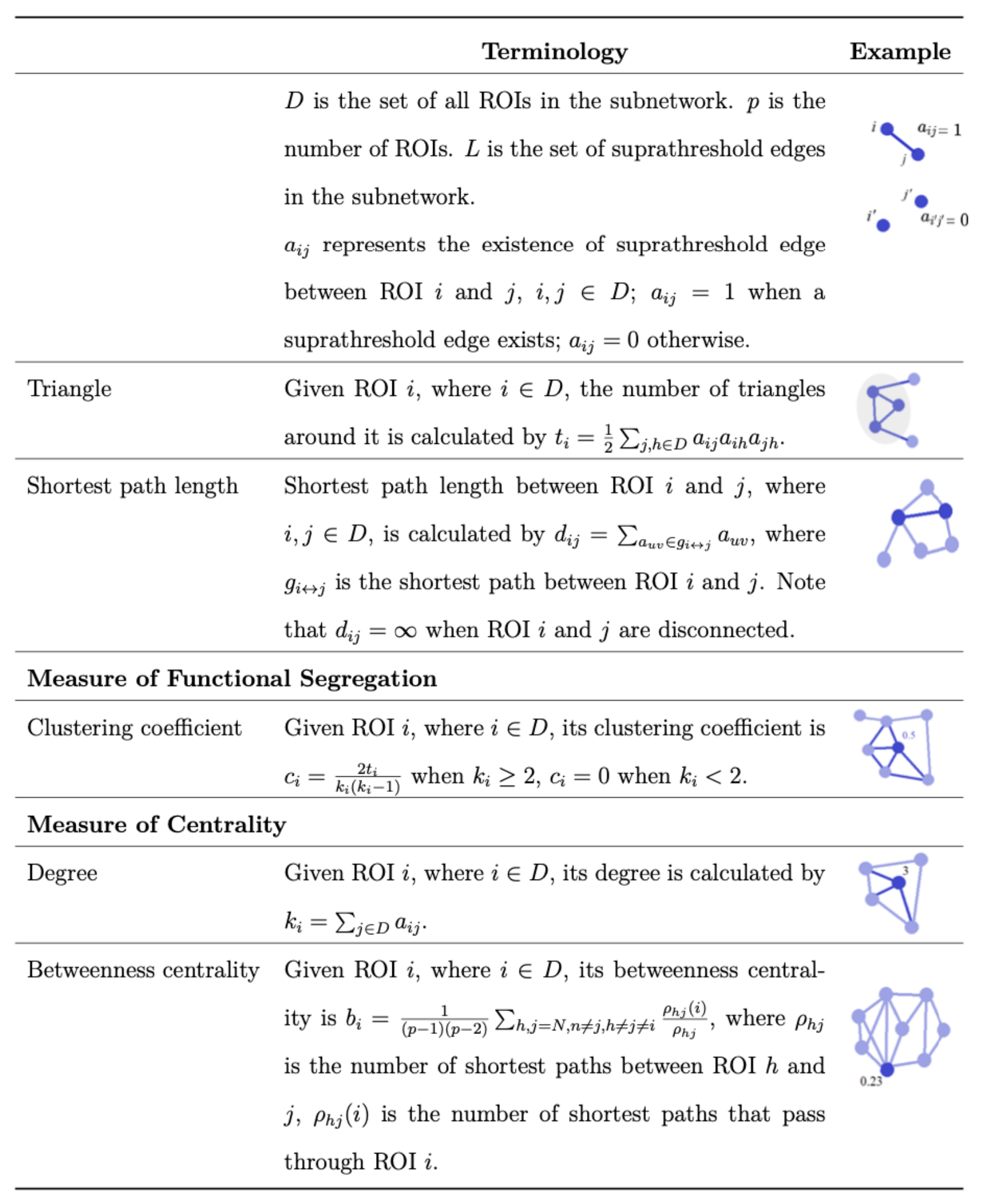}
  \caption{\label{fig: network_measures}Terminology and examples of basic network measures, clustering coefficient, degree, and betweenness.}
\end{figure*}

We used complex network measures to quantify the topological properties of the subnetwork \citep{Rubinov2010} (Figure~\ref{fig: network_measures}). We focused on the  functional segregation and centrality measures. Functional segregation is the ability of processing a certain task to occur within densely interconnected ROIs \citep{Rubinov2010}. The clustering coefficient is a basic measure of functional segregation based on the number of triangles around a given ROI, where each triangle is formed by the given ROI and two other distinct ROIs, and the three edges connecting them. The clustering coefficient of the ROI quantifies how well connected this ROI's neighbors are, which is equivalent to the number of triangles around the ROI divided by the number of edges that could possibly exist between the ROI and its neighbors. The ROI with a clustering coefficient that is close or equal to $1$ implies that the other ROIs in this subnetwork cluster around it. On the other hand, central nodes or hubs are the ROIs with central placement in the overall subnetwork structure. They play an important role in communication and integration in the subnetwork \citep{Vandenheuvel2013}. Degree, a simple measure of centrality, is defined as the number of edges directly linked to a given ROI. The larger the degree is, the more central the ROI is. Betweenness is another measure of centrality defined at a given ROI as the fraction of all shortest paths in the subnetwork, which are paths with the smallest numbers of edges between two ROIs in the subnetwork, passing through the ROI. An ROI with high betweenness is regarded as a bridge connecting the other ROIs in the subnetwork. 

\subsection{Comparison with standard lesion symptom mapping}

We used three traditional mass univariate methods: voxel-, region-, and connectivity-based lesion symptom mapping (VLSM, RLSM, CSLM). Whole brain V- and RLSM was used to identify brain damage associated with aphasia type (anomic or Broca's). VLSM shows the statistical likelihood that damage to a given voxel is associated with aphasia type group membership, where each voxel in each patient is binarily demarcated as either damaged or undamaged \citep{Bates2003}. RLSM differs from VLSM in that instead of using binary voxel-wise values, it uses the percent of voxels damaged within each ROI as the predictor of aphasia type. This sacrifices spatial specificity while providing the advantage of analyzing the effects of damage over an entire region without requiring overlapping damage at level of an individual voxel. We conducted RLSM using the AICHA ROIs. We then conducted CLSM \citep{Gleichgerrcht2017} using resting-state functional connectivity based on the AICHA atlas, including all left-to-left, left-to-right, and right-to-right connections in the analysis. Only voxels (or regions for RLSM) where at least 5 patients had damage were considered, based on the minimum overlap recommendation of ~10\% of the patient sample \citep{Baldo2022}. All tests were two-tailed, with $\alpha = 0.05$, and significance was determined via permutation testing, 
where stability of $p$-value were tested in increments of 1000 permutations, ranging from 1,000 permutation to 10,000 permutations. 

On top of whole-brain analysis, we also restricted the analysis to the 'dorsal stream' areas, i.e. frontoparietal and superior temporal areas that are involved in form-to-articulation during speech \citep{Fridriksson2016}. These areas would be hypothesized to be especially disrupted in individuals with Broca's aphasia who struggle with many aspects of speech production compared to the relatively mild anomic cases where the individuals just have occasional word-finding difficulties. We included the AICHA ROIs corresponding to supramarginal gyrus, primary sensory and motor cortices, inferior frontal gyrus (Broca's area), superior temporal gyrus, and rolandic operculum. This allowed us to restrict the \# of connections while also allowing us to use a one-tailed analysis since we specifically hypothesized these connections would be associated with Broca's aphasia. It is worth noting that we also tried the alternate analysis, using a different set of language regions that might be implicated in anomic aphasia more than Broca's, but this did not reveal any significant results. 
This is likely because anomic aphasia as a behavioral syndrome may be caused by deficits at various functional levels within the language production system (conceptual, lexical, semantic, phonological, for example), so that similar surface behavior may result from different patterns of neural damage. In addition, in our own sample anomic aphasia was 'less severe' than Broca's aphasia, on average, which would also make the detection of areas specifically related to the anomic group more difficult.

\section{Results}

All simulation studies and data analyses in this study were performed with MATLAB R2020a. 

\begin{table}[t!]
	\centering
	\begin{threeparttable}
	\begin{tabular}{l l l l l l}
		\toprule
		&  & \textbf{NBS} & \multicolumn{3}{c}{\textbf{Mass Univariate Testing}}\\
		\cmidrule{4-6} $(p, C_r)$ & & & \textbf{Bonf.} & \textbf{Holm's} & \textbf{FDR}\\
		\hline
		{\textbf{\emph{Study 1}}}&\textbf{TPR} & 0.7278 & 0.2538 & 0.2538 & 0.5896 \\
		(20, 10) &\textbf{FPR} & 0.0089 & 0.0002 & 0.0002 & 0.0016 \\
		\hline
		{\textbf{\emph{Study 2}}}&\textbf{TPR} & 0.9557 & 0.5670 & 0.5670 & 0.9431\\
		(40, 10) &\textbf{FPR} & 0.0103 & 0.0001 & 0.0001 & 0.0024\\
		\hline
		{\textbf{\emph{Study 3}}}&\textbf{TPR} & 0.9317 & 0.5687 & 0.5687 & 0.9270 \\
		(40, 5) &\textbf{FPR} &  0.0104 &	0.0001 & 0.0001 & 0.0012 \\
		\bottomrule
	\end{tabular}
	\caption{The average TPR and FPR of the NBS with threshold 2.5 and mass univariate testing with multiple comparison, in 5,000 simulations.}
	\label{Tab: average TPR and FPR}
	\end{threeparttable}
\end{table}

\begin{table}[b!]
	\centering
	\begin{threeparttable}
	\begin{tabular}{c c c c}
		\toprule
		\textbf{Threshold} & \textbf{Number of} & \textbf{Number of} & \textbf{$p$-value} \\
		& \textbf{ connections} & \textbf{ ROIs} & \\
		\hline
		3.5	&	271	&	72&	0.0078 \\
		
		4	&	85 &	47&		0.0032 \\
		
		4.5	&	34	& 	27&		0.0010 \\
		
		5	&	9  &   7&		0.0010 \\
		\bottomrule
	\end{tabular}
	\caption{The subnetwork identified by the NBS method with threshold 3.5, 4, 4.5 and 5.} 
	\label{Tab:componet results}
	\end{threeparttable}
\end{table}

\subsection{Simulation results} 

Table \ref{Tab: average TPR and FPR} summarizes results of the three simulation studies. In Study 1, the NBS method has the largest TPR and FPR, while the mass univariate testing with FDR detects a desirable proportion of contrast edges and contains a small number of false discoveries. In Study 2, as the network size expands, the TPRs by the NBS and mass univariate testing methods increase, whereas the FPRs increase by the NBS and FDR methods and decrease by the Bonferroni and Holm's Bonferroni methods. 
In Study 3, when fewer contrast edges are placed in either of the two groups, the TPRs by the NBS and FDR methods decrease and the FPRs by the NBS and the two Bonferroni corrections 
stay similar to Study 2. 

In summary, compared with mass univariate testing, the NBS method detects small group differences well under various network sizes and proportions of contrast edges. For the mass univariate testing, the FDR has the highest power and a favorable FPR; Bonferroni and Holm's Bonferroni corrections are highly conservative in detecting contrast edges. Additionally, we find that the computation speed is mainly affected by network size. 

\subsection{NBS results on POLAR rs-fMRI} 

\begin{figure*}[t!]
	\centering
	\includegraphics[width=\linewidth]{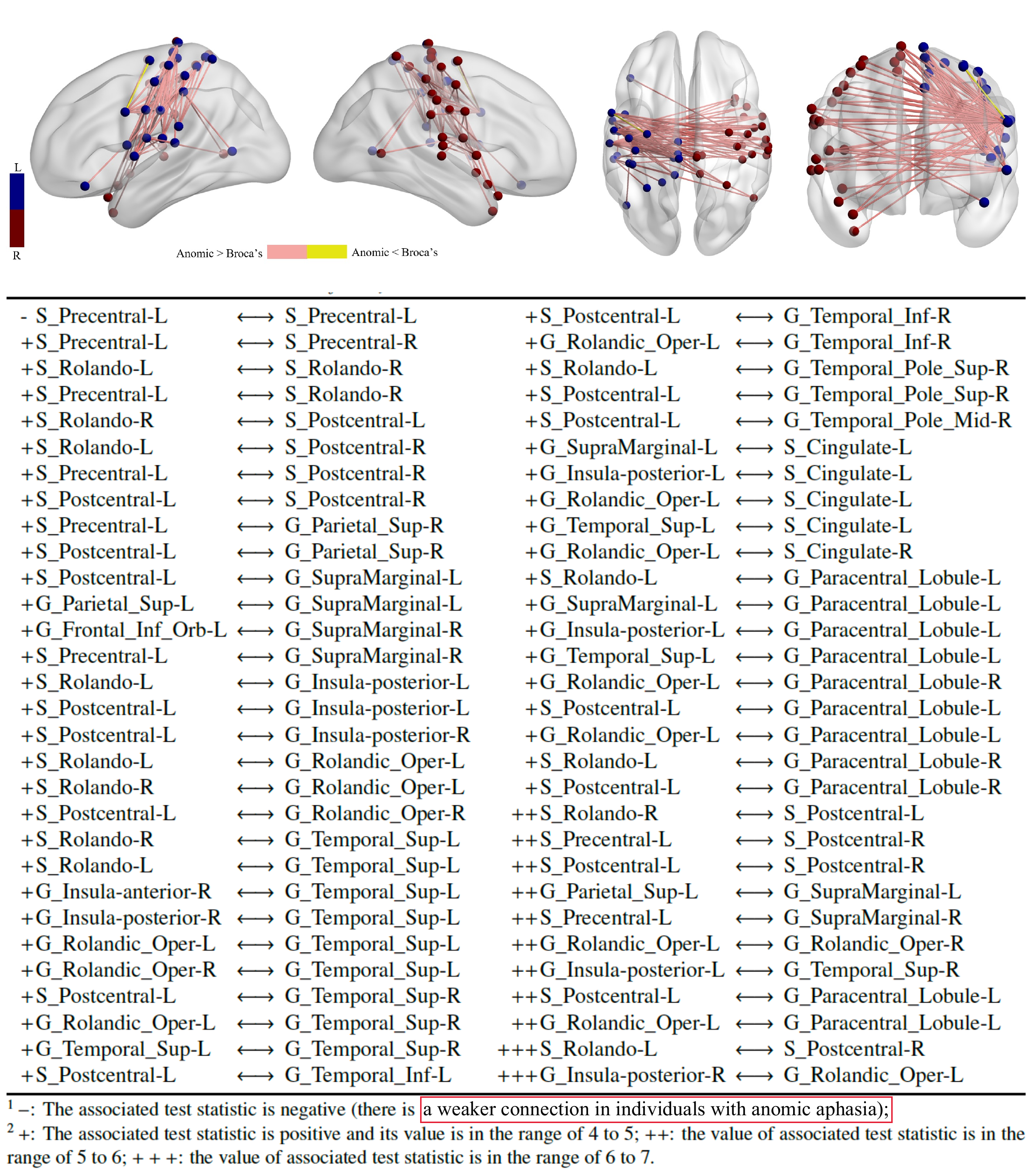}
	\caption{Connections in the subnetwork identified by the NBS method with threshold 4. Figures at the top were plotted by the BrainNet Viewer \citep{Xia2013}). From the sagittal view, we found some left ROIs are connected ipsilaterally, while all right ROIs are connected to left ROIs.}
	\label{fig: Subnetwork in whole brain}
\end{figure*}

\begin{figure*}[t!]
	\includegraphics[width=\textwidth]{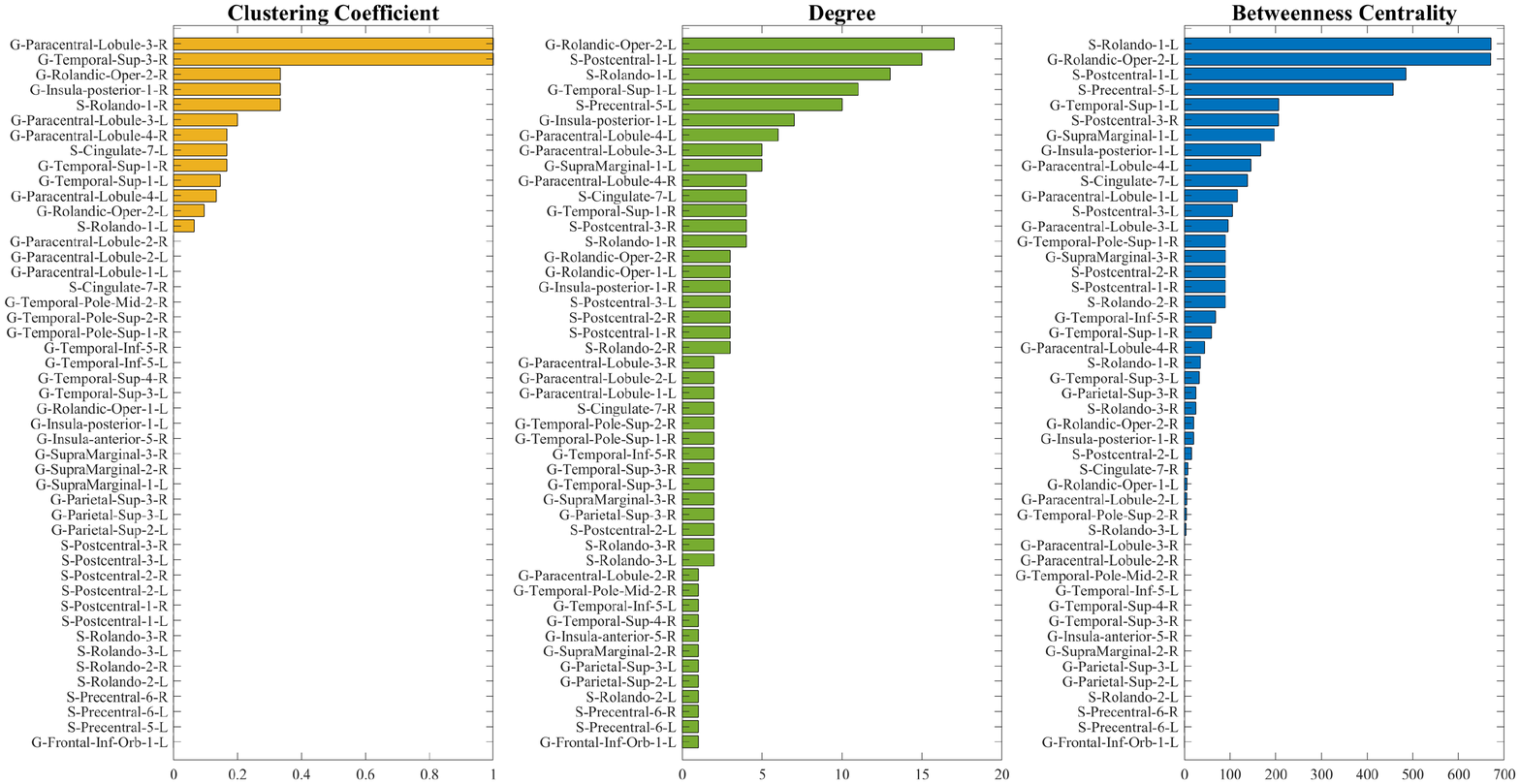}
	\caption{Complex network measures for the identified subnetwork}
	\label{fig:complex network measures of the subnetwrok in both hemispheres}
\end{figure*}

Applying the NBS method to the POLAR rs-fMRI, we identified a subnetwork distinguishing the anomic and Broca's groups with respect to each of the thresholds 3, 3.5, 4 and 5. As the threshold increased, the subnetowrk size and the number of ROIs contained in the subnetwork decreased (Table \ref{Tab:componet results}). Only one subnetwork was identified when the threshold was set to 4. Subsequent results all pertain to this subnetwork. Figure \ref{fig: Subnetwork in whole brain} shows connections comprising this subnetwork. Except for one connection within the left precentral sulcus (between middle and superior precentral sulcus), the majority of connections in the networks from the group with anomic aphasia were stronger than the networks from the people with Broca's aphasia. The subnetwork was nearly bilaterally symmetric and primarily involved the ROIs in the premotor, primary motor, primary auditory, and primary sensory cortices.  A few ROIs in the left Broca's area and cingulate cortex were also included in this subnetwork. In comparison, mass univariate testing with multiple comparison also identified several connections that were significantly different between the two aphasia groups. By comparing the results in the two methods, we found that the majority of significant connections declared by mass univariate testing existed in the subnetwork identified by the NBS with threshold 4 and above. When the threshold in the NBS was less than 4, a large amount of connections in the subnetwork were not detected by mass univariate testing.

For topological properties of the subnetwork identified by NBS with threshold 4, the clustering coefficient, degree, and betweenness as described in the previous section were computed (Figure \ref{fig:complex network measures of the subnetwrok in both hemispheres}). The clustering coefficient of the ROIs in the superior temporal gyrus (STG) and auditory cortex (Aud) in the right hemisphere reached the maximum value of 1. The clustering coefficient of the ROIs in the left STG and Aud was also higher than other ROIs in this subnetwork. Regarding the centrality, we found there were more than 10 connections linked to the ROIs in the left premotor, primary sensory, and primary motor cortices and thus these ROIs had a larger degree than others. Furthermore, the ROIs in the left premotor, primary motor cortices and the right supramarginal gyrus exhibited a high betweenness.

\begin{figure*}[t!]
  \includegraphics[width = 1\linewidth]{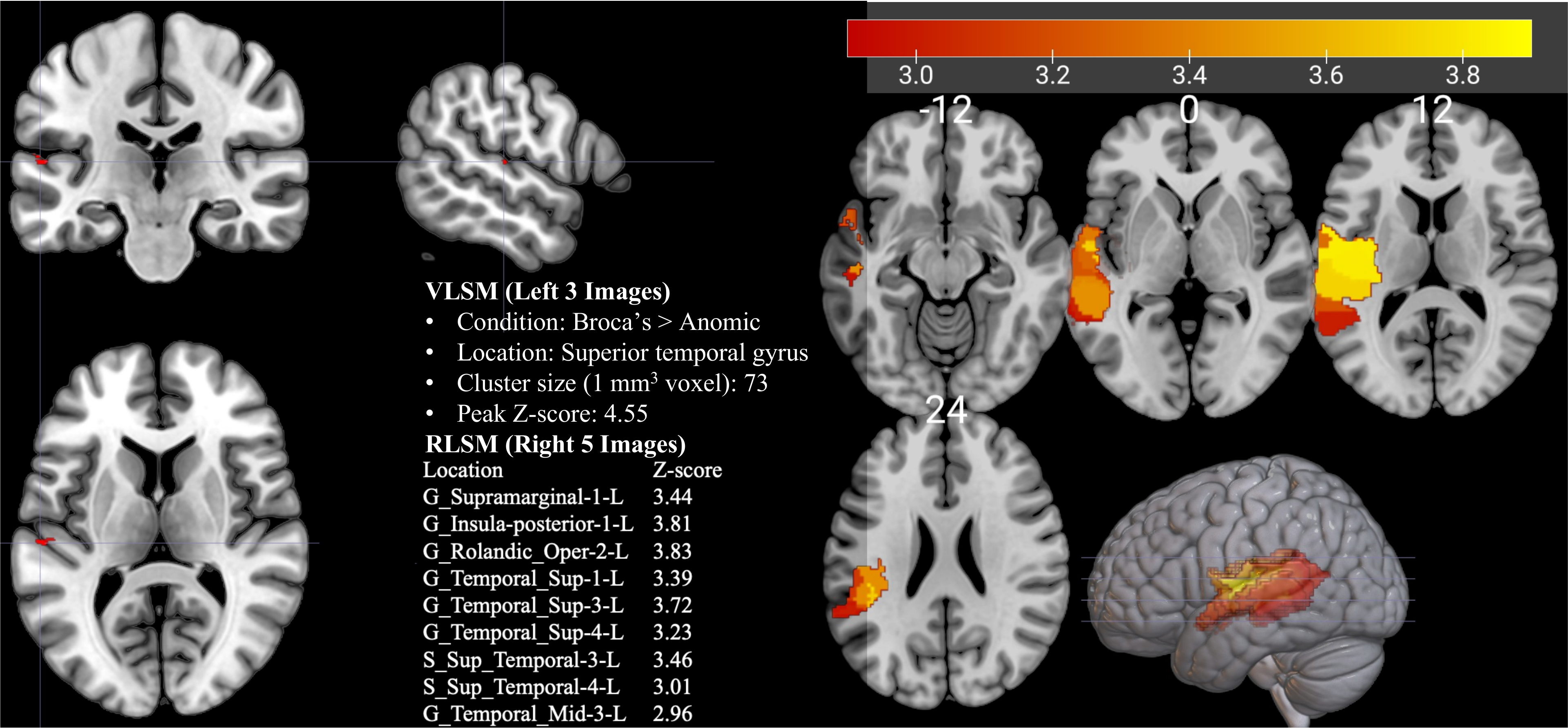}
  \caption{\label{fig: lsm}Whole brain VLSM revealed a small cluster of voxels in the left superior temporal gyrus where damage was significantly associated with Broca’s aphasia group membership (peak z = 4.55, p < .001). RLSM showed that percent of voxels damaged within 9 left hemisphere AICHA regions was associated with Broca’s aphasia group membership, including portions of STG, MTG, and insula.}.
\end{figure*}

\subsection{Standard lesion symptom mapping (LSM) results}

Whole-brain VLSM revealed a small cluster of voxels in the left superior temporal gyrus where damage was significantly associated with Broca’s aphasia group membership (peak $z = 4.55$, $p < .001$; Figure~\ref{fig: lsm}). RLSM showed that percent of voxels damaged within 9 AICHA ROIs on the left hemisphere was associated with Broca’s aphasia group membership, including portions of STG, MTG, and insula (Figure~\ref{fig: lsm}). On the other hand, no connections were significant in the CLSM analysis. Using the set of AICHA ROIs hypothesized to be more involved in Broca's aphasia than anomic, only the connection from G\_Rolandic\_oper\_1\--R to G\_Temporal\_Sup\_1\--L survived and is significantly weaker in the Broca group ($z = -4.12$). Restricting the analysis to just interhemispheric (left to right) connections revealed two additional connections (left posterior insula to right superior temporal gyrus, and left superior temporal gyrus to right superior temporal gyrus). Overall, significant results remained unchanged as we increased the \# of permutations from 1,000 up to 10,000. Critical values only changed slightly depending on \# of permutations. 

\section{Discussion}
This study identified functional subnetworks distinguishing anomic and Broca's aphasia. The subnetwork identified through the NBS method under threshold 4 is located in the premotor, primary motor, primary auditory, and prime sensory cortices in both hemispheres. Existing studies have shown that the premotor and primary motor cortices contribute to a higher level of motor function \citep{Knierim2000}. For instance, premotor and primary motor cortices are involved in encoding complex patterns of motor actions, selecting appropriate motor plans, or controlling a series of movements. In severe Broca’s aphasia, the damage in Broca’s area, lower primary motor cortex and surrounding tissue can leave the patient with no speech output or only stereotyped output, but relatively intact comprehension \citep{Naeser1989}. The weaker connections in motor, sensory and auditory cortices for Broca's aphasia explain the typical articulatory difficulties in Broca's aphasia, likely due to impaired motor movement planning, which is rarely observed in anomic aphasia. Another finding is that the identified subnetwork involves both hemispheres, with only a small part of the subnetwork limited to Broca's area and cingulate cortex of the left hemisphere. Though not exclusively \citep{Matchin2020}, Broca's area in the left hemisphere has long been associated with syntactic processing both during production and comprehension \citep{Price2012}. Consistent with this, the reduced connections between ROIs in left-hemisphere Broca's area, as a part of the subnetwork, may account for differences in sentence and grammatical language production between anomic and Broca's aphasia. 

In the conceptual framework for the analysis of word-finding difficulties proposed by \citet{Rohrer2007}, word retrieval failure in anomic aphasia is due to deficits in vocabulary, since speakers with anomic aphasia can use approximate expressions to substitute for the object they want to say, and their comprehension of the meaning of the object is preserved. Our finding shows that the connection within the left-hemisphere precentral sulcus is reduced in anomic aphasia, which suggests another potential substrate for this type of aphasia. 

Using measures of functional segregation, we observed intensive interactions around the ROIs located in the bilateral STG and Aud. Many neuroimaging studies related to language processing claim that the functional organization in the bilateral STG is involved in spoken word recognition, particularly its phonological stage \citep{Hickok2009, Binder2000, Binder1994}. The right hemisphere alone is suggested to be capable of good auditory comprehension at the word level \citep{McGlone1984, Zaidel1985}. Although both anomic and Broca’s aphasia are characterized by relatively preserved comprehension, repetition performance at the word and sentence levels is typically more affected in Broca’s aphasia, and reduced interactions between STG and auditory cortex may well underlie this reduced performance. 

Using measures of centrality, the degree of the ROIs in the left primary motor, premotor, and primary sensory cortices shows that more connections linking to them than other ROIs in the subnetwork. The left primary motor, premotor, and primary sensory cortices are mainly involved in the articulation network, contributing to speech production over language-related tasks, and to a lesser extent to nonspeech oral-motor movements \citep{Basilakos2018, Friedemann2006}. These left dominant regions are considered higher-order areas in processing for speech and motor function, which extract and process the information acquired from primary sensory receptive areas. 

Furthermore, the results of betweenness suggest that the left premotor and primary motor cortices play a crucial role in information flow and overall communication efficiency in the subnetwork and comprise the bridge connecting to other ROIs. The result also reveals the importance of the right supramarginal gyrus in this subnetwork. This finding is unexpected, since the known functions of the right supramarginal gyrus for healthy and stroke-affected participants primarily include proprioception, particularly for upper limbs \citep{Ben-Shabat2015}, as well as overcoming emotional egocentricity bias and controlling empathy towards other people \citep{Silani2013}. Its role in language processing has not been described in detail, although \citet{Hartwigsen2010} reported that activation in right-hemisphere SMG was likely associated with stroke survivors’ attempts at compensating for impaired performance on a phonological judgement task.

For comparison, we also ran standard VLSM, RLSM, and CLSM, where CLSM is most closely related to NBS in its connection-based inference approach. But no connection survived in the whole-brain CLSM on the same AICHA ROIs for NBS, where we used all connections (including L-L, L-R, and R-R). 
We then restricted to the set of regions hypothesized to be more involved in Broca's aphasia than anomic. Only one connection G\_Rolandic\_oper\_1\--R to G\_Temporal\_Sup\_1\--L survived for Broca $>$ Anomic, and none for Anomic $>$ Broca. These results indicate that with traditional CLSM, the correction for multiple comparisons is severe due to the large number of possible connections. Restricting the number regions increases the power, but even then the penalty can be severe, unless the analysis is restricted to very few a priori regions. NBS can provide much greater sensitivity and allow exploration of large networks and their features.

To develop a better understanding of the deficits in language processing streams from a network perspective, further research could investigate comparisons and distinctions between other types of aphasia. One of the advantages of the NBS method is controlling the FWER at a subnetwork level; while the standard mass univariate testing approach is controlling at individual connection independently, which ignores the interactions among ROIs. If suprathreshold edges form a subnetwork, the NBS method provides substantially greater power \citep{Zalesky2010}. Nevertheless, if ROIs are not connected by suprathreshold edges and thus cannot form a subnetwork, the NBS method will fail to make any decision. Under a network framework, once the NBS identifies any subnetwork, it would allow us to study their network topology and make straightforward interpretations. These findings may be somewhat limited by not taking into account the effect sizes and significance of the differences. For instance, after a threshold is applied, the weight for suprathreshold edges would all set to 1. In future investigations, it may be possible to use weighted edges and take account of variation in differences. On the other hand, the critical factor affecting the subnetwork size and determining whether suprathreshold edges would form a subnetwork is the threshold in the NBS method. 
Under different thresholds, new subnetwork(s) with fewer or more functional connections may be identified, and the $p$-value of the subnetwork(s) may change and result in a loss of significance. To address the concern of threshold, we examined a range of thresholds in the NBS method and incorporated the baseline approach in our analysis. The functional connections are significant across both the NBS method with a range of thresholds and the baseline approach, which implies notable differences in functional connectivity between anomic and Broca's aphasia. For future studies, the NBS method can be combined with a data-driven threshold selection or threshold-free approach. This would yield a subnetwork with a more precise boundary for network comparison. 

\section{Conclusion}
In this study, by comparing resting-state functional connectivity between stroke survivors with anomic and Broca's aphasia, a distinct subnetwork was identified. The subnetwork is mainly located in the premotor, primary motor, primary auditory, and prime sensory cortices bilaterally. The majority of connections among these cortices are weaker in Broca’s aphasia than anomic aphasia. By examining network properties of this subnetwork, we found (1) the brain regions located in the bilateral STG and Aud show intensive interaction; (2) primary motor, premotor and primary sensory cortices in the left hemisphere exhibit high centrality. These findings suggest that the disruptions in interactions among motor, auditory and sensory cortices cause the differences in a resting baseline between anomic and Broca's aphasia. In further studies, we can explore multiscale network models to overcome the arbitrary thresholding issue \citep{Wang2022}, as well as task-based functional connectivity. 

\section{Acknowledgments}
\indent The authors would like to thank Dr. Grigori Yourganov at Clemson University, and Drs. Roger Newman-Norlund and Lorelei Phillip Johnson at the University of South Carolina for providing access of the POLAR dataset used in this study. 

\section{Declarations}
\subsection{Author Contributions} 
\noindent Conceptualization (XZ, YW, JF). Study design and data acquisition (JF). Statistical analysis (XZ). Interpretation of results (XZ, YW, DdO, RD). Writing (All). 
\subsection{Funding Sources}
\noindent RD: R01DC017162 and R01DC01716202S1. \\
\noindent JF: NIH
R21-DC014170 and P50-DC014664.

\subsection{Compliance with Ethical Standards} The research was approved by the Institutional Review Board (IRB) at the University of South Carolina.
\subsection{Conflict of Interest and Disclosure}
None.

\bibliographystyle{apacite}
\bibliography{references}
\end{document}